\documentclass[article, amsmath, amsthm, amssymb, accents, aps,]{revtex4-2}
\usepackage{amsmath, amsthm, amssymb, accents}
\newcommand{\myvect}[1]{\accentset{\rightharpoonup}{#1}}
\usepackage[pdftex]{graphicx}
\begin{document}
\title{Could a gravity inversion exist within the Hollow Earth of Legendary's Monsterverse?} 
\author{R. Steven Millward}
\email{smillward@sd148.org}
\affiliation{Math and Science Departement Faculty Member\\
Grace High School, Grace, Idaho \\
Mathematics and Physics Department Adjunct Faculty Member \\
Idaho State University, Pocatello, Idaho}
\date{\today}
\begin{abstract}
One of the most popular movie franchises of late is Legendary's Monsterverse as is evidenced by the gross earnings of the series recently surpassing the two and a half billion dollar mark with the release of $\it{Godzilla X Kong}$ \cite{Monster}.  As is typical with many movies of this genre, in order to enjoy them one must suspend their disbelief when it comes to the laws of physics.  While there are a plethora of examples of violations of basic principles (the square-cube law being the prime example \cite{Square}) the idea of a "gravity inversion" occurring inside of the  "Hollow Earth" is among the most debated as well as the most relevant to the storylines of the recent movies.  The intent of this paper is to show that, while typical scientific arguments refuting the possibility of such an inversion are definitely completely correct, a slight modification of the conditions assumed to exist inside the planet in these arguments may allow for the inversion to actually occur.
\end{abstract}
\maketitle

The movies of Legendary's "Monsterverse" have, as a central tenet, that the interior of the Earth is configured differently from the typical explanation given by scientists.  The basic premise is that there are regions inside the Earth that are "hollow".  This is known in the movies as the "Hollow Earth Theory" (HET).  (It should be noted that a real theory of a hollow Earth was put forward in the 17th century by Edmund Halley which was subsequently proven incorrect. \cite{Halley})   Most serious scientists would scoff at the suggestion of a hollow Earth for a variety of reasons.  I do not intend to defend the hollow Earth theory from these legitimate, well-founded arguments, all of which I personally believe in.  However, one of the predictions of the HET is that at some critical point between the outer "shell" of the Earths surface and the inner "core" there exists a "gravity inversion".  At points closer to the core than this critical point, the direction of gravity still points toward the center of the Earth as usual.  However, at points closer to the shell than the critical point, the local direction of gravity is directed away from the center of the Earth.  This allows the various creatures in the movies not only to walk on the outside of the core in the usual fashion but also to walk equally well on the inner surface of the shell.   I would simply like to focus on one particular argument usually used to dismiss the gravity inversion.

For most purposes, Newtonian theory is sufficient to describe the gravitional configuration of the Earth.  Keeping this in mind we note three well-known facts about Newtonian gravity \cite{Physics}.  First, Newtonian gravity is linear.  This means that if we have multiple mass configurations and want to calculate the gravitational field due to all of them we can simply find the potential of each piece and then add the individual potentials together before taking the negative gradient to find the total field.  Second, it is well known that the gravitational field outside a sphere of radius $a$ and density $\rho_1$ is given by
$$
\myvect{g}=-{{4\pi G \rho_1 a^3} \over 3 r^2} \hat{r}
$$
Third is the fact that at all points inside a spherical shell, because of symmetry, the gravitational field is zero.  Therefore, if the hollow Earth is modeled as a spherical shell surrounding a solid sphere with air of negligible density in between, the superposition of the two gravitational fields is just that of the inner sphere.  Thus the gravity inversion that the movies portray is incorrect.  My intent is to show that if the shell of the hollow earth is not symmetric then there can exist at least one point where the local direction of gravitational acceleration switches directions, as suggested in the movies.

We start by assuming that the cross-section of the hollow Earth is represented by Figure 1.
\begin{figure} [h]
\begin{center}
\includegraphics[scale=0.70]{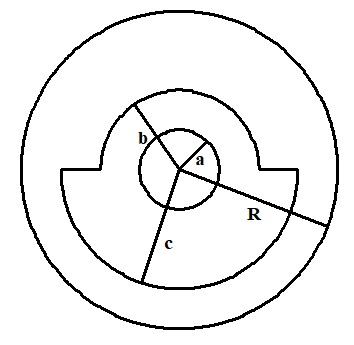}
\caption{}
\end{center}
\end{figure}
The core of radius $a$ in the center has a constant mass density of $\rho_1$ throughout.  Both the thick top shell and thin bottom shell have a constant mass density of $\rho_2$.  Note that this setup still has azimuthal symmetry but not latitudinal symmetry.  Using the usual spherical coordinates, we would like to calculate the potential "directly above" the core (at $\theta=0$ and $a < |z| < b$) and "directly below" the core (at $\theta=\pi$ and $a < |z| < c$).  We can then calculate the gravitational field using the equation $\myvect{g}=-\myvect{\nabla}{V_t}$.  We will do this by calculating the potentials of the core, upper shell and lower shell individually, taking the negative gradient, and then combining them together and evaluating whether a critical point exists at which there is an inversion.  This is represented as $V_t=V_c + V_{us} + V_{ls}$.

We have already expressed the field due to the core. The core potential at a distance $|z| > a$ above or below the core is given by $V_c=-{{4\pi G \rho_1 a^3} \over |z|}$.  We now must find the potential due to both shells.  The gravitational potential of a mass configuration of density $\rho(r')$ is given by the integral
$$
V=-G\int\limits_{V'}{{\rho(r')} \over {|\myvect{r} - \myvect{r}'|}} dV',
$$
where the density in the shells is $\rho_2$ which is constant and $\myvect{r}$ is the vector to the point the field is to be measured and $\myvect{r}'$ is the integration variable used to characterize the location of the mass distribution.  In spherical coordinates, $|\myvect{r} - \myvect{r}'|$ is given by,
$$
|\myvect{r} - \myvect{r}'|=\sqrt{r^2 + {r'}^2 - 2rr'(\sin{\theta}\sin{\theta '}\cos[\phi - \phi '] + \cos{\theta} \cos{\theta '}}.
$$
Since we are calculating the field value on what we have labeled the $z$-axis, $\theta = 0$ or $\theta = \pi$ and $r=z$.  With these values, we can rewrite the denominator of the integral as,
$$
|\myvect{r} - \myvect{r}'|=\sqrt{{r'}^2 \mp 2z\cos{\theta '} r' + z^2},
$$
where the $\mp$ sign is for the field at points above the core and below the core, respectively.  Note that there is no $\phi '$ dependence left, so that part of the volume integral can be calculated and yields a multiplicative factor of $2\pi$ to the $G$.  The limits on the $r'$ and $\theta '$ integrals are different for each shell.  Thus the integrals to be calculated are given by,
$$
V_{us} = -2\pi G \rho_2 \int\limits_{b}^{{R}} \int\limits_{0}^{{\pi \over 2}} {{{r'}^2 \sin{\theta '} d\theta ' dr'} \over \sqrt{{r'}^2 \mp 2z\cos{\theta '} r' + z^2}}
\hspace{0.3in}
V_{ls} = -2\pi G \rho_2 \int\limits_{c}^{{R}} \int\limits_{{\pi \over 2}}^{\pi} {{{r'}^2 \sin{\theta '} d\theta ' dr'} \over \sqrt{{r'}^2 \mp 2z\cos{\theta '} r' + z^2}}
$$

Now, if we do the $\theta '$ integral first, we can treat $r'$ as a constant and make a coordinate transformation.  Defining $s^2=r'^2\mp2z\cos{\theta '}r' + z^2$ we see that $2s ds = \pm 2z\sin{\theta '}r' d\theta '$.  The $0$, ${\pi \over 2}$ and $\pi$ limits of integration get replaced by $r' \mp z$, $\sqrt{r'^2 + z^2}$ and $r' \pm z$, respectively.  With this transformation, the integral simplifies dramatically to
$$
V_{us}=-2\pi G \rho_2 \int\limits_{b}^{R} \int\limits_{r' \mp z}^{\sqrt{r'^2 + z^2}}\pm {r' \over z}ds dr'.
$$
Both the $s$ and $r'$ integral are now easily solvable.  Upon performing the integrations, the potential of the upper shell is found to take the form
$$
V_{us}=\mp 2 \pi G \rho_2 \left({{{(R^2 + z^2)}^{3/2} - {(b^2 + z^2)}^{3/2} + b^3 - R^3} \over {3z}} \pm {R^2 \over 2} \mp {b^2 \over 2}\right)
$$
An identical solution scheme allows us to solve for the potential of the lower shell, with the only differences being the corresponding limits of integration.  When the dust settles the potential of the lower shell is found to be,
$$
V_{ls}=\mp 2 \pi G \rho_2 \left({{{-(R^2 + z^2)}^{3/2} + {(c^2 + z^2)}^{3/2} - c^3 + R^3} \over {3z}} \pm {R^2 \over 2} \mp {c^2 \over 2}\right)
$$
Combining these two potentials with that of the core yields the total gravitational potential of the asymmetric hollow Earth setup.  It is
$$
V_t=-{{4\pi G \rho_1 a^3} \over |z|} + 2 \pi G \rho_2\left(\pm {{{(b^2 + z^2)}^{3/2} - {(c^2 + z^2)}^{3/2} + c^3 - b^3} \over 3z} + {{(b^2 + c^2)} \over 2} - R^2\right)
$$
We are now at a point where we can find the gravitational field along the $z$-axis between the core and shells.  Taking the negative gradient of the total potential yields,
$$
\myvect{g}=-2 \pi G\left(\pm{{2 \rho_1 a^3} \over z^2} + {{\rho_2(b^3 - c^3)} \over {3z^2}} \pm \rho_2\left[\sqrt{b^2 + z^2} - \sqrt{c^2 + z^2} + {{\sqrt{{(c^2 + z^2)}^3}} \over 3z^2} - {{\sqrt{{(b^2 + z^2)}^3}} \over 3z^2}\right]\right)\hat{k}
$$
Finally, getting a common denominator of $3z^2$ and simplifying gives the final form for the gravitational field given below.
$$
\myvect{g} = {{2\pi G} \over {3z^2}}\left(\mp 6\rho_1 a^3 + \rho_2[c^3 - b^3] \mp \rho_2\left[(2z^2 - b^2)\sqrt{b^2 + z^2} - (2z^2 - c^2)\sqrt{c^2 + z^2}\right]\right)\hat{k}
$$
Note that, if $b=c$, we have the symmetric shell which contributes nothing to the field and we just have the field of the core, pointing down toward the core when $\theta=0$ and up toward the core when $\theta = \pi$.  Recall that these two values of $\theta$ correspond to the $\mp$ signs in the formula.

In order for an inversion to exist, the expression in $\left( \right)$'s must be continuous and go to zero at a valid value of $z$ (between $a$ and $b$ for the point above the core and between $a$ and $c$ for the point below the core).  Finding a solution analytically is at least very difficult if not impossible.  However, we can numerically look for a value of $z$ with the properties we want.  Before doing so, we must give values for the parameters $\rho_1$, $\rho_2$, $a$, $b$, $c$.  If we assume the core is similar to the core of the Earth and the shell is similair to the crust of the Earth, then $\rho_1 \approx 3\rho_2$.  The movies do not specify exact thicknesses of the core or shells.  This leaves the parameters corresponding to the core and shell thicknesses undetermined.  Without exact values to plug in, we simply look to see if there is a combination of $a$, $b$ and $c$ for which the inversion occurs halfway between the core and shell, as specified in some of the movie novelizations.

For the $\theta = \pi$ location, we find no combination of parameters that yield a root for the expression in parentheses given above.  For the $\theta=0$ location we find a number of different combination of parameters that will give an inversion at a legitimate value of $z$.  We show one such combination in the figure below.
\begin{figure} [h]
\begin{center}
\includegraphics[scale=0.50]{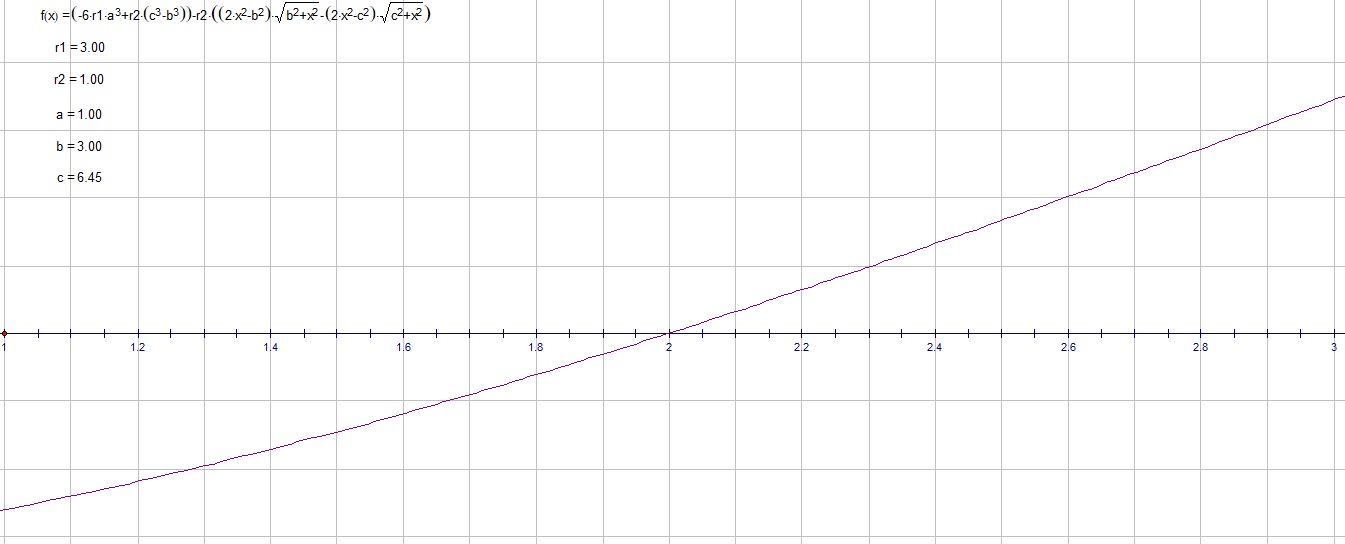}
\end{center}
\end{figure}

In the figure, $\rho_1=3$, $\rho_2=1$, $a=1$, $b=3$, $c=6.45$ and the inversion point is found at $z=2$.  

In conclusion, a number of issues should be addressed.  First, there is no guarantee that the above configuration is stable.  The core is most likely not at an equilibrium point and will be drawn toward the thick shell above it if no other forces prevent it from doing so.  Second, I have not investigated the physical ramifications of the particular combinations of parameters used to find the above inversion solution.  It may well be that this combination is untenable for the Earth.  Third, the discontinuity in the thickness of the shell is unphysical and only used because I was able to find a solution using it.  A gradual increase of shell thickness as a linear function of latitudinal angle was investigated but eventually abandoned due to the difficulty in integrating to find the potential of the shell.  Fourth, it is stressed that this inversion point is truly a point.  Any other off-axis regions between the core and shell may or may not have this inversion behavior.  Finally, this discovery in no way makes the HET of the movies a plausible theory in real life.  As mentioned above, numerous arguments \cite{Hollow} can be found dealing with temperature, pressure, energy source, etc. that effectively rules it out.  However, in a very particular situation, it has been shown that the idea of a gravity inversion is at least somewhat plausible.  For the configuration explored, one could, in principle, jump off the core, pass through the inversion point, land on the inside of the shell and be stuck there with their notion of up and down having switched. 


\newpage
\begin{thebibliography} {1}
\bibitem{Monster}  The Numbers, Nash Information Services, Retrieved on August 5th, 2024 from https://www.the-numbers.com/movies/franchise/MonsterVerse
\bibitem{Square} Sketchplanations, Retrieved on August 5th, 2024 from https://sketchplanations.com/the-square-cube-law
\bibitem{Halley} Kollerstrom, N. (1992). The Hollow World of Edmond Halley. Journal for the History of Astronomy, 23(3), 185-192. https://doi.org/10.1177/002182869202300304
\bibitem{Physics} Bauer, W. and Westfall, Gary, University Physics, McGraw Hill Publishing, 2011
\bibitem{Hollow} Ward, C., GODZILLA VS. KONG'S 'HOLLOW EARTH' WHERE GIANT MONSTERS LIVE IS BOGUS, RIGHT? THE SCIENCE BEHIND THE FICTION, Syfy Website, Retrieved on August 5th, 2024 from https://www.syfy.com/syfy-wire/godzilla-vs-kong-hollow-earth-science-true-false-real-fake
\end{thebibliography}
\end{document}